# Most borrowed is most cited? Library loan statistics as a proxy for monograph selection in citation indexes

Álvaro Cabezas-Clavijo[1], Nicolás Robinson-García[1], Daniel Torres-Salinas[2], Evaristo Jiménez-Contreras[1], Thomas Mikulka[3], Christian Gumpenberger[3], Ambros Wernisch[3] & Juan Gorraiz[3]

[1] *acabezasclavijo@gmail.com, {elrobin, evaristo} @ugr.es*
EC3: Evaluación de la Ciencia y de la Comunicación Científica, Departamento de Información y Comunicación, Universidad de Granada, Campus de Cartuja s/n E-18071 Granada (Spain)

[2] *torressalinas@gmail.com*
EC3: Evaluación de la Ciencia y de la Comunicación Científica, Centro de Investigación Médica Aplicada, Universidad de Navarra, Pamplona (Spain)

[3] *{christian.gumpenberger, ambros.wernisch, thomas.mikulka, juan.gorraiz} @univie.ac.at*
University of Vienna, Vienna University Library, Boltzmanngasse 5, A-1090 Vienna (Austria)

**Abstract**

This study aims to analyse whether library loan statistics can be used as a measure of monograph use and as a selection criterion for inclusion in citation indexes. For this, we conducted an exploratory study based on loan data (1000 most borrowed monographs) from two non-Anglo-Saxon European university libraries (Granada and Vienna) with strong social sciences and humanities components. Loans to scientists only were also analysed at the University of Vienna. Furthermore, citation counts for the 100 most borrowed scientific monographs (SM) and textbooks or manuals (MTB) were retrieved from Web of Science and Google Scholar. The results show considerable similarities in both libraries: the percentage of loans for books in national languages represents almost 96% of the total share and SM accounts only for 10%–13%. When considering loans to scientists only, the percentage of English books increases to 30%; the percentage of SM loans also increases (~ 80%). Furthermore, we found no significant correlations between loans and citations. Since loan statistics are currently insufficient for measuring the use of monographs, their suggested use as an applicable selection criterion for book citation indexes is not yet feasible. Data improvement and aggregation at different levels is a challenge for modern libraries in order to enable the exploitation of this invaluable information source for scientometric purposes.

**Conference Topics**

Topic 1: Scientometric Indicators
Topic 2: Old and New Data Sources for Scientometric Studies

**Introduction**
Bibliometric indicators have increasingly been used for research assessment purposes since the 1970s. Among other uses, they have been applied to implement reward mechanisms within academia, exceeding their original purpose which was to serve as an aid for journal selection in university libraries. In this sense, bibliometric studies have rested primarily on two basic metrics: journals' impact factors and citations of papers. However, unlike journal articles for which new indicators have been developed in the last few years (SJR, SNIP, Eigenscore), monographs, an important scholarly communication channel for the humanities and social sciences fields, have been left behind. The absence of books in the main databases with bibliometric data has led evaluation agencies to consider monographs a minor scientific product. This has resulted, in many of these fields, in the devaluation of monographs. In fact, as shown in the UK, many researchers have shifted from books to journal articles as their preferred dissemination product due to the pressure exerted by national evaluations (Research Information Network, 2009).

There has been little exploration of usage indicators within the scientometric community as a proxy to measure the use and impact of academic materials. The advent of the digital format in academia has brought the development of new tools, such as journal hubs or repositories, which produce new use-derived indicators. These metrics represent a potential opportunity for applying alternative evaluation methods, aiming to complement or even replace the traditional bibliometric indicators based on citations. Projects such as COUNTER (Counting Online Usage of Networked Electronic Resources) or MESUR (Metrics from Scholarly Usage of Resources) have worked on this line of work, developing the necessary frameworks and standards to achieve such goals.

Likewise, the adoption of so-called web 2.0 tools by researchers has added a new dimension in which usage indicators can also be applied to measure the impact of research, not only within the academic community but also in society at large. The main characteristics of these indicators, known as Altmetrics (Priem et al, 2010), are: 1) they work at an item level, and 2) they can be obtained in real time. However, they also have many shortcomings, such as the evanescence of data or complexity in terms of apprehending their real meaning. The adoption of these indicators by journals such as PLoS One, publishing houses such as Nature Publishing Group, or major databases such as Scopus, highlight the level of acceptance they are gaining within the scientific community.

Despite the availability of e-books on scientific platforms and in library catalogues, these usage indicators do not perform well in this context such that they might be considered valuable alternatives for research assessment. However, there are other usage indicators that could be used instead. Library loans may capture the impact of books in ways that e-usage and Altmetrics cannot. This is particularly interesting in the fields of the social sciences and humanities, where monographs still play a significant role. This approach is not new as Price (1963) remarked that the "amount of usage provides a reasonable measure of the scientific importance of a journal or a man's work". Following this line of thought, loans could not only be used as proxies of dissemination, but also of relevance or importance, detecting books that may be considered top tier research outcomes for inclusion in citation indexes. This is as a relevant issue as evaluation agencies tend to assess based only the content of these indexes.

The launch of the Thomson Reuters' Book Citation Index (hereafter, BKCI) in 2010 introduced new elements of study for the bibliometric community. Beyond the citations that each book or book chapter gather from other citation indexes, this product also allows the analysis of publishers or citation patterns within book-oriented areas (Gorraiz et al, 2013; Leydesdorff & Felt, 2012; Torres-Salinas et al, 2012). Thomson Reuters states that "there is a need to select those publications that will most likely contain significant scholarship" and that "priority is given to books and book series that have relatively greater citation impact" (Testa, 2012). Such statements suggest that the mere indexing of monographs within this database is a sign of quality commensurate with that for journals and therefore it also suggests that solid methodology must be developed to ensure that the materials to be included in the future are chosen fairly. Currently, apart from certain bibliographic requirements and the citation impact, the company does not specify further criteria for inclusion.

Based on the premise that library statistics are an invaluable and underutilized source of information for assessment purposes, we explore in this paper the extent to which library loans might be used as a proxy for the measurement of monograph use. Furthermore, we test the feasibility of using library loans as a possible selection criterion for monographs in citation indexes.

**Theoretical Background**

The influence of monographs in the social sciences, especially in the arts and humanities, as the main communication channel between scholars has traditionally led to serious shortcomings when adopting bibliometric methodologies to analyse and assess research activity in these areas. Indeed, not only they lack sound and consolidated bibliometric measures, but their nature poses many limitations that must be overcome in order to apply the correct methodological procedure when employing them. In this sense, there are three main issues that should be resolved as a prior step before any kind of methodological approach is taken: the definition of monograph types, the establishment of significant proxies of quality, and addressing differences across disciplines.

Regarding the first issue, monographs are extremely heterogeneous as content of a different nature can be found in them. According to Testa (2011), in the BKCI the following book formats are considered scholarly: dissertations, textbooks, books in series, reprinted/reissued content, translations and non-English content, biographies and reference books. However, this classification is of no use for bibliometric purposes as the ambiguity in the definition of each of these formats prevents us from establishing a relation between citation or usage patterns and document types as is done with journal articles (reviews, letters, notes, research articles, etc.). In relation to this, Torres-Salinas and Moed (2009) also point out the lack of book typologies as a shortcoming when assessing monographs and suggest three possible criteria: a) authorship of the monograph (authored versus edited works), b) research intensity (books primarily for teaching versus books primarily for research), and c) research focus (books for a specialized scientific audience versus books for broader audiences). However, such classifications cannot be found in databases or library catalogues, forcing us either to make no distinction, or use an erroneous classification which may lead to the wrong conclusions, or perform a manual classification. An alternative approach would be to classify them according to the traditional book categories described in library catalogues (i.e., handbook, manual, report and research work) and analyse them separately, deciding the level of research intensity after the analysis has been undertaken.

The second issue to take into account is the proxy chosen as a performance indicator when evaluating scholarly monographs. Until the launch of the BKCI, it remained extremely difficult to assess books using citation data for large sets of records; in fact, few studies can be found using large citation data sets regarding the evaluation of books (Gorraiz et al, 2012; Leydesdorff & Felt, 2012; Torres-Salinas et al, 2013). What is more, the few studies that use citations as a proxy warn that the patterns observed have many peculiarities that must be taken into account when they are used. One of them has to do with the aging of citations, as monographs seem to need much wider citation windows (Cronin, Snyder & Atkins, 1997) than two to five years, as is common with journal articles. This may be a good explanation for the findings described by Torres-Salinas et al (2012), who observed that more than 70% of the books and book chapters indexed since 2005 in the BKCI remained uncited.

In order to solve not only these shortcomings but also the issue regarding data availability, other proxies have been suggested in the literature for evaluating monographs based on: a) library holdings, such as the number of catalogue entries per book title in WorldCat® (Torres-Salinas & Moed, 2009), library bindings (Linmans 2010), or even introducing an indicator of perceived cultural benefit (White et al, 2009); b) document delivery requests (Gorraiz & Schlögl, 2006); c) publishers' prestige (Giménez-Toledo, Tejada-Artigas & Mañana-Rodríguez, 2012); d) book reviews (Zuccala & van Leeuwen, 2011). However, none of these has been adopted unanimously by the bibliometric community, mainly because obtaining the data is difficult and time consuming.

Finally, the last issue that needs to be resolved has to do with the different practices observed across disciplines. This is also observed when analysing journal articles, but it may be even more acute with monographs as these are most commonly used in the humanities and social sciences in which practices are more fragmented and there is a strong national factor biasing researchers' behaviour (Hicks, 1999). The role played by monographs is especially important in these fields as they form one of the main channels for scholarly communication (Hicks, 2004; Research Information Network, 2009, 2011; Williams et al, 2009).

**Data & Methodology**
We conducted a pilot study to test the extent to which library loans could be used as a proxy to measure monographs' relevance within the scientific community. Such analysis was performed in two non-Anglo-Saxon European university libraries: the library of the University of Granada (Spanish-speaking) and the Vienna University Library (German-speaking). Both of them are universities with several centuries of history and both libraries are universal with strong social sciences and humanities components.

*Briefbackground of the institutions (structure) and description of their loan systems:*
A) Granada:
One of the historic universities of Spain, the University of Granada was founded in 1531. Despite its encyclopaedic character, it is a university with strong social sciences and humanities components, these being areas to which 47.6% of the research staff are affiliated (University of Granada, 2011). Its library system comprises 21 libraries which are located within faculties, institutes and research centres, providing services to more than 80,000 students, 3,650 researchers and 2,000 technical and administrative staff. According to the last library report (University of Granada, nd), there are 1,042,575 monographs. The integrated library system developed by Innovative was established in 2001 in the library premises; the loan system was centralized by means of Millenium software and affords a number of different reports concerning loans and renewals of monographs and other materials.

B) Vienna
Founded in 1365, the Vienna University Library is the oldest university library in the German-speaking countries;[1] it is also the largest library in Austria with an inventory of over 6.8 million books. It comprises a main central lending library (2.6 million volumes) and 40 specialist libraries, providing researchers, teachers and students with specialist literature on almost all specific academic subjects. It provides services to approximately 91,000 students and a university staff of 9,400 employees, of whom 6,700 are academic. For some disciplines, such as medicine, veterinary medicine, economics, agriculture and technical subjects, there are additional universities with corresponding university libraries in the city. Since the winter semester of 1986, loans have been managed electronically; in 1989, an electronic catalogue was also introduced. In 1999, the Aleph integrated library system replaced the previous software. In 2010, around 4 million book loans and 65,000 active borrowers were counted.

*Data retrieval and processing*
Data were gathered from the Universities of Vienna and Granada library systems in December 2012 regarding loans from, respectively, 2001 and 2000 onwards. For every monograph copy we recorded the following fields: book title, author, location, publisher, country and year of edition, language, year of acquisition by the library, ISBN, number of loans, and number of renewals.

For the University of Granada, only copies of bibliographic material with more than 50 loans were recorded. Afterwards, loan data for every title were aggregated, regardless of their publication year or publisher. Books with the same title but different authors were also detected and considered separately. For the Vienna sample, loan data were analysed at the level of bibliographical records. Thus, all copies of a certain edition were automatically aggregated, but different editions of one work were covered separately. Also, Granada distinguished book types according to the library's classification. For this study, we decided to use three main book types:

- REF = reference books,such as dictionaries, etc.
- MTB = manuals, textbooks, handbooks, etc.
- SM= scientific monographs

---
[1] http://bibliothek.univie.ac.at/english/about_us.html

These bibliographic types are assigned by librarians at the institution. However, some titles were assigned to two different types (MTB and SM). For those materials, all the copies were recoded as MTB. We also coded as MTB those monographs with the following words in their titles for Spanish and English language books: Course, Encyclopedia, Foundations, Introduction, Methods, Principles, Treaty, Grammar, Atlas, Compendium, Handbook and Textbook. Additionally, after carefully checking every SM title, some other books were recoded as MTB. Finally, books with Law, Code or Dictionary in the title were recoded as reference books (REF). As Vienna lacked a classification of materials, this differentiation was manually performed by two librarians in line with the criteria used by the University of Granada and described above. The error ratio was less than 5%.

For this first pilot analysis, we retrieved the 1000 most borrowed monographs by all types of users since the loan system at each library was implemented (1999–2000 for Vienna, 2001 for Granada). At the University of Vienna and in view of the initial results, it was also possible to collect data on the 1000 monographs most borrowed exclusively by scientists and without considering those from the Faculty of Law library. We then performed analyses and comparisons between all samples at three different levels: book types, language and publisher. Furthermore, for the 100 first most borrowed scientific monographs (SM) and the 100 first most borrowed manuals and textbooks (MTB) citation counts were performed:1) in Web of Science using the "Cited Reference Search", and 2) in Google Scholar using the Publish or Perish software (Harzing, 2007), which calculates various bibliometric indicators based on the data retrieved from this database. All citation data were manually disambiguated and aggregated in both data sources.

**Results**

*A) Loans by document type*

The results for both universities are summarized in Table 1. As already mentioned above, the analyses were performed in both universities at edition level and not at title level. In the sample from Vienna (Top 1000, all users) only five books (4 MTB and 1 REF) had the same ISBN and approximately 12% were related to more than one edition (thereof ~ 60% REF). In the second sample (Top 1000, only scientists),only one book (SM/MTB) had the same ISBN twice and only 1.1% of the titles had multiple editions (thereof ~ 60% REF). For the University of Vienna, 119 monographs (~ 10%) comprised both sample 1 and sample 2 (~ 36% SM, ~ 36% REF, ~ 28% MTB). The Pearson correlation between loans to all users and loans to scientists only was rather low (about 0.20).

**Table 1. Loans by document type for the three samples**

| | TYPE | TITLES | % | LOANS | % | LOANS/TITLE | MAX | MIN |
|---|---|---|---|---|---|---|---|---|
| **Granada** | MTB | 706 | 70.6 | 290943 | 81.3 | 412 | 3922 | 110 |
| | SM | 245 | 24.5 | 49465 | 13.8 | 202 | 844 | 110 |
| | REF | 49 | 4.9 | 17300 | 4.8 | 353 | 1751 | 113 |
| | TOTAL | 1000 | 100.0 | 357708 | 100.0 | 358 | 3922 | 110 |
| **Vienna** | MTB | 334 | 33.4 | 166895 | 27.5 | 500 | 2372 | 207 |
| | SM | 200 | 20.0 | 60121 | 9.9 | 301 | 1087 | 206 |
| | REF | 466 | 46.6 | 380651 | 62.6 | 817 | 7090 | 207 |
| | TOTAL | 1000 | 100.0 | 607667 | 100.0 | 608 | 7090 | 206 |
| **Vienna Only Scientists** | MTB | 162 | 16.2 | 1690 | 15.9 | 10 | 30 | 8 |
| | SM | 782 | 78.2 | 8199 | 77.2 | 10 | 29 | 8 |
| | REF | 56 | 5.6 | 738 | 6.9 | 13 | 33 | 8 |
| | TOTAL | 1000 | 100.0 | 10627 | 100.0 | 10627 | 33 | 8 |

In the sample from Granada, more than 70% of the most borrowed books (706 titles) were found to be manuals, textbooks and handbooks (MTB), this percentage increasing to 81.3% when taking loans into account. Loans per title for the top MTB were twice the loans per title for top scientific monographs (SM). This category accounts for 24.5% of the loans, while REF books (mainly dictionaries) are just 4.9% of the loans for the Spanish library (Table 1).

The results show a very similar number of scientific monographs in both universities (20% vs. 24%) when considering loans to all users. The differences between the other document types (MTB and REF) can be explained by the large number of books related to law at the University of Vienna (almost 50%), most probably due to their high prices. When considering the number of loans, the rate for scientific monographs drops to 10% in agreement with a lower loan frequency in comparison with the other document types, MTB and REF (see "loans per title" in Table 1). When considering loans to scientists only, the amount of SM grows abruptly to 78%.

*B) Loans by language*

The results for all three samples are represented in Table 2. Granada and Vienna show very similar results when we consider the language distribution of the top 1000 borrowed books. Almost identical percentage values were reported in both universities: about 94% of the 1000 most borrowed titles are in each country's language, Spanish and German. The percentage of loans for books in national languages is even higher (96%). Despite being the scientific "lingua franca", English is not popular within these academic communities as in both universities the percentage of books in English stays at around 5%. For the only scientists sample in Vienna, this percentage grows to more than 30%, showing an important difference in comparison to the all users sample.

**Table 2. Loans by language for the three samples**

| | LANGUAGE | TITLES | % TITLES | LOANS | % LOANS | LOANS/TITLE | MAX | MIN |
|---|---|---|---|---|---|---|---|---|
| **Granada** | Spanish | 938 | 93.8 | 343935 | 96.1 | 367 | 3922 | 110 |
| | English | 47 | 4.7 | 10750 | 3.0 | 229 | 675 | 133 |
| | Multilingual | 7 | 0.7 | 1344 | 0.4 | 192 | 431 | 137 |
| | French | 3 | 0.3 | 938 | 0.3 | 313 | 649 | 111 |
| | German | 3 | 0.3 | 420 | 0.1 | 140 | 157 | 114 |
| | Italian | 2 | 0.2 | 321 | 0.1 | 161 | 191 | 130 |
| | TOTAL | 1000 | 100.0 | 357708 | 100.0 | 358 | 3922 | 110 |
| **Vienna** | German | 945 | 94.5 | 582559 | 95.9 | 616 | 7090 | 206 |
| | English | 55 | 5.5 | 25108 | 4.1 | 457 | 1221 | 206 |
| | TOTAL | 1000 | 100.0 | 607667 | 100.0 | 608 | 7090 | 206 |
| **Vienna Only Scientists** | German | 653 | 65.3 | 7093 | 66.7 | 11 | 33 | 8 |
| | English | 318 | 31.8 | 3250 | 30.6 | 10 | 30 | 8 |
| | French | 13 | 1.3 | 109 | 1.0 | 8 | 13 | 8 |
| | Italian | 7 | 0.7 | 68 | 0.6 | 10 | 10 | 8 |
| | Serbian | 3 | 0.3 | 45 | 0.4 | 15 | 20 | 10 |
| | Spanish | 1 | 0.1 | 9 | 0.1 | 9 | 9 | 9 |
| | Lithuanian | 1 | 0.1 | 9 | 0.1 | 9 | 9 | 9 |
| | Croatian | 1 | 0.1 | 17 | 0.2 | 17 | 17 | 17 |
| | Rumanian | 1 | 0.1 | 8 | 0.1 | 8 | 8 | 8 |
| | Czech | 1 | 0.1 | 10 | 0.1 | 10 | 10 | 10 |
| | Hungarian | 1 | 0.1 | 9 | 0.1 | 9 | 9 | 9 |
| | TOTAL | 1000 | 100.0 | 10627 | 100.0 | 11 | 33 | 8 |

*C) Loans by publisher*

Tables 3–5 show the top publishers according to the number of titles (or editions) respectively for all three samples.

The alternation of national (Ariel, Pirámide, Manz, Facultas WUV, etc.) and international publishers (McGraw Hill, Prentice, Springer, Pearson, etc.) is similar in both distributions (see Tables 3 and 4). The highest concentration reported for Vienna (only two publishers, Manz and Facultas.WUV account for almost half of the most borrowed books while in Granada ten publishers are needed to surpass 50% of loans) is explained by the predominant role of Manz as the Austrian publisher for law and other reference texts related to law (see also results by document type). In contrast, the names and proportions of the international publishers in both rankings are quite different. For example, McGraw-Hill, top in Granada, does not appear in the top 20 for Vienna, and Springer, top in Vienna, is not present in Granada's top ranking.

**Table 3. Distribution of the top 20 publishers in the Granada sample**

| PUBLISHER | TITLES | LOANS | % LOANS | LOANS/TITLE |
|---|---|---|---|---|
| McGraw-Hill | 105 | 47690 | 13.3 | 454 |
| Ariel | 39 | 12927 | 3.6 | 331 |
| Pirámide | 37 | 12736 | 3.6 | 344 |
| Prentice Hall | 37 | 17220 | 4.8 | 465 |
| Alianza Editorial | 35 | 10511 | 2.9 | 300 |
| Edit. Médica Panamericana | 32 | 15701 | 4.4 | 491 |
| Síntesis | 27 | 15762 | 4.4 | 584 |
| Tecnos | 26 | 14353 | 4.0 | 552 |
| Masson | 25 | 6805 | 1.9 | 272 |
| Pearson Education | 23 | 10722 | 3.0 | 466 |
| Tirant lo Blanch | 22 | 17298 | 4.8 | 786 |
| Omega | 18 | 9580 | 2.7 | 532 |
| Elsevier | 18 | 7641 | 2.1 | 425 |
| Comares | 18 | 4738 | 1.3 | 263 |
| Thomson | 17 | 5167 | 1.4 | 304 |
| Universidad de Granada | 15 | 4980 | 1.4 | 332 |
| Reverté | 14 | 5736 | 1.6 | 410 |
| Oxford University Press | 13 | 3991 | 1.1 | 307 |
| Addison Wesley | 13 | 6291 | 1.8 | 484 |
| Difusión | 11 | 3117 | 0.9 | 283 |
| Akal | 11 | 2828 | 0.8 | 257 |
| Cátedra | 10 | 1730 | 0.5 | 173 |

**Table 4. Distribution of the top 20 publishers for the Vienna sample (all users)**

| PUBLISHER | TITLES | LOANS | % LOANS | LOANS/TITLE |
|---|---|---|---|---|
| Manz | 174 | 177505 | 29.2 | 1020 |
| Facultas.WUV | 132 | 93853 | 15.4 | 711 |
| LexisNexis-Verl. ARD Orac | 123 | 66405 | 10.9 | 540 |
| Springer | 65 | 47741 | 7.9 | 734 |
| Pearson Prentice-Hall | 38 | 19689 | 3.2 | 518 |
| Linde | 35 | 17944 | 3.0 | 513 |
| Beltz | 23 | 14982 | 2.5 | 651 |
| VS Verl. Für Sozialwiss. | 19 | 6847 | 1.1 | 360 |
| Thieme | 16 | 6661 | 1.1 | 416 |
| Böhlau | 16 | 9520 | 1.6 | 595 |
| Hogrefe | 16 | 7996 | 1.3 | 500 |
| Oldenbourg | 14 | 8702 | 1.4 | 622 |
| SpektrumAkad. Verl. | 13 | 6332 | 1.0 | 487 |
| Westdt. Verl. | 13 | 4363 | 0.7 | 336 |
| Huber | 12 | 4687 | 0.8 | 391 |
| Verl. Österreich | 11 | 7379 | 1.2 | 671 |
| Wiley-VCH | 11 | 4336 | 0.7 | 394 |
| Leske + Budrich | 9 | 2927 | 0.5 | 325 |
| UVK-Verl.-Ges. | 9 | 3386 | 0.6 | 376 |
| Elsevier, SpektrumAkad. Verl. | 9 | 3174 | 0.5 | 353 |

When comparing both samples, the publisher distribution is considerably less concentrated for only scientists (~340 publishers) than for all users (~140 publishers for Vienna and ~240 for Granada). Other particularities for the Vienna loans to scientists only (see Table 5) are the appearance of new publisher names, amongst them many foreign university press companies, and the homogeneity of the number of loans per title for all top publishers which fluctuate between nine and 13 loans.

**Table 5. Top publishers with more than 10 titles for the University of Vienna sample, only scientists**

| PUBLISHER | TITLES | LOANS | % LOANS | LOANS/TITLE |
|---|---|---|---|---|
| Suhrkamp | 46 | 492 | 4.6 | 11 |
| Böhlau | 38 | 423 | 4.0 | 11 |
| Oxbow Books | 32 | 295 | 2.8 | 9 |
| Manz | 28 | 365 | 3.4 | 13 |
| Cambridge Univ. Press | 26 | 253 | 2.4 | 10 |
| Campus-Verl. | 25 | 296 | 2.8 | 12 |
| Springer | 24 | 302 | 2.8 | 13 |
| Oxford Univ. Press | 22 | 214 | 2.0 | 10 |
| Beck | 21 | 230 | 2.2 | 11 |
| Routledge | 20 | 216 | 2.0 | 11 |
| Fink | 20 | 203 | 1.9 | 10 |
| Facultas.WUV | 18 | 193 | 1.8 | 11 |
| VS Verl. Für Sozialwiss. | 15 | 145 | 1.4 | 10 |
| Transcript | 13 | 126 | 1.2 | 10 |
| de Gruyter | 11 | 110 | 1.0 | 10 |
| Tempus | 10 | 92 | 0.9 | 9 |
| Metzler | 10 | 118 | 1.1 | 12 |

**Correlations between loans and citations**

The mean number of loans for the 100 most borrowed scientific monographs (SM) and the 100 most borrowed manuals (MTB) was 771.6 (Granada) and 640 (Vienna) respectively, showing for both samples a much higher number of loans for those books coded as MTB. Regarding citations, the results show a mean value of 171.5 and 260.7 respectively for Google Scholar, and 25.4 and 53.7 respectively when using Web of Science to retrieve citations. The median of citations was 36 and 18 respectively for the GS sample and only two and 7.5 respectively for the Web of Science sample. When comparing SM and MTB, the differences in loans and citations were statistically significant (CI=95%, $p< 0.05$), MTB median values being much higher than the SM values, with the only exception being Google Scholar citations for the Vienna sample (see Table 6). It is also worth mentioning the sizeable differences between numbers of citations in both databases, with Google Scholar being more exhaustive.

**Table 6. Loans and citations analyses for the 100 most borrowed SM and MTB titles**

|  | MTB | SM | TOTAL |
|---|---|---|---|
| **GRANADA** | | | |
| loans | 1245,6 ± 575,7 (1061,5) | 297,7 ± 132,4 (254,5) | 771,6 ± 632,0 (698,0) |
| citations_GS | 173,9 ± 387,7 (58,5) | 169,1 ± 533,1 (25,0) | 171,5 ± 465,0 (36,0) |
| citations_WOS | 22,8 ± 66,4 (4,0) | 28,1 ± 167,3 (1,0) | 25,4 ± 127,0 (2,0) |
| **VIENNA** | | | |
| loans | 908,1 ± 330,8 (841,5) | 371,9 ± 110,8 (336,5) | 640,0 ± 364,4 (580,0) |
| citations_GS | 244,7 ± 805,1 (7,0) | 276,6 ± 1475,5 (23,0) | 260,7 ± 1185,7 (18,0) |
| citations_WOS | 62,5 ± 200,2 (12,5) | 44,9 ± 138,9 (6,0) | 53,7 ± 172,1 (7,5) |

Mann-Whitney Test: CI=95%; $p<0.05$. Results are reported as Mean ± Standard Deviation (median)

Finally, we performed a correlation analysis by means of the Spearman coefficient (Rho) to test the extent to which loans and citations gathered by both methods were similar. The results show that for the Granada case, there is no correlation at all between loans and citations, regardless of the citation source used and the type of monograph (Table 7). Only correlations between citations gathered from both sources were found to be statistically significant. It is worth noting that this value is higher for scientific monographs (0.765; 0.481 for Vienna) than for handbooks (0.577; 0.466 for Vienna). A statistically significant correlation (0.310) between loans and citations as measured by Google Scholar was detected for MTB books in the Vienna sample. However, this correlation is so weak that is not appropriate to infer any kind of consistent finding regarding loans and citations. Also, the correlation between citations regarding both databases for MTB and SM books was very weak (less than 0.5).

**Table 7. Spearman correlation coefficients for loans, citations and book type**

|  |  | MTB | | SM | |
|---|---|---|---|---|---|
|  |  | citations_GS | citations_WOS | citations_GS | citations_WOS |
| **GRANADA** | | | | | |
| loans | Rho | 0.135 | 0.081 | 0.099 | 0.115 |
|  | Sig. | 0.181 | 0.421 | 0.328 | 0.254 |
| citations_GS | Rho |  | 0.577* |  | 0.765* |
|  | Sig. |  | 0.000 |  | 0.000 |
| **VIENNA** | | | | | |
| loans | Rho | 0.310* | 0.189 | 0.032 | 0.060 |
|  | Sig. | 0.002 | 0.059 | 0.751 | 0.550 |
| citations_GS | Rho |  | 0.466* |  | 0.481* |
|  | Sig. |  | 0.000 |  | 0.000 |

*Correlation is statistically significant at 0.01 level (2-tailed)

**Discussion and concluding remarks**

In this paper we analyse whether library loans might be used as a proxy for the measurement of monograph use and the feasibility of using library loans as a possible selection criterion for including monographs in citation indexes. To this end, we conducted an exploratory study analysing loan data and citation data from two university libraries which represent two non-Anglo-Saxon academic communities, a Spanish-speaking community and a German-speaking community.

Methodologically, this has not been an easy process. A number of technical factors need to be considered before taking loans as a valid measure for the useof monographs and subsequently as a selection criterion for book citation indexes: the publication year is not the acquisition year, extensions, loan times or loan counts, differentiation of the user types and materials, different counts in different libraries, multiple editions and copies, etc. Also, the loan time can differ between universities and types of users and materials. Finally, some books could be classified as not for loan due to several reasons, so no data could be gathered for them. It is also worth mentioning the presence of library departments which loan books that may not be within the general automated library system.

As shown in our study, different approaches can be taken, such as measuring numbers of copies, editions or titles of monographs. Also, the aggregation of counts from different editions and translations should be considered, as a number of the most "popular" books happen to be translations of Anglo-Saxon monographs. Counting only the translations could miss the academic impact of a book as a whole. One additional technical difficulty is the detection of citations referring to different books where the title coincides in various languages, as happens for titles such as Economia (Economy) or Biologia (Biology) in Spanish, Portuguese and Italian. Usually handbooks and manuals have a broad coverage, so the titles are very short (one or two words in many cases), which makes it impossible to split citations for every language.

Most of the top books were manuals, handbooks and textbooks, or reference books such as those for law and dictionaries.This is understandable as the main users of university libraries are students. The results also show that scientific monographs in both universities account for 20%–25% of the most borrowed books when considering loans for all users. However, this percentage increases to 78% when analysing loan patterns for scientists only. This fact points to the need to differentiate between types of users to assess more precisely the reliability of loans as a usage indicator for monographs.

A second conclusion is that both academic communities (Vienna and Granada) prefer to borrow books in their respective languages, regardless of the original language of the publication. The outstanding percentage (around 95% for Spanish and German) of loans for publications in national languages could be explained by the type of users, mainly students. When assessing the loan behaviour of scientists only (solely for the Vienna sample), we have found that they are more likely to borrow English monographs than the other user groups. This is not surprising as these monographs are expected to convey more specialized information which could justify the lack of a translation of such books. However, for the Vienna scientists, German is also the preferred language when looking up information in scientific monographs, with more than 65% of the most borrowed books being in German.

The rankings of publishers according to the most borrowed books also show the presence of both international publishers and local well-known publishing houses. Both of these tend to distribute scientific monographs and handbooks with a broad orientation and these are the materials most borrowed from academic libraries.

We also have found important differences in the number of citations retrieved using Web of Science and Google Scholar. Google Scholar has gathered more citations than Web of Science for most of the samples when considering the median number of citations. The exception is the Vienna sample for MTB where Web of Science retrieves a higher number of mentions with respect to those monographs. Standard deviations for citation statistics suggest that within the most borrowed books, highly cited books coexist with other monographs that are not noticed by the scientific community.

Regarding citations relating to the most borrowed books, it is important to mention the lack of correlation between these two variables in our study for the Granada sample and the very weak correlation for Vienna MTB books when using Google Scholar. The fact that the books cover a broad range of topics, mainly in languages other than English and of a general nature, may be important for the interpretation of this finding. We also have to consider the different citation behaviours in each discipline and the aging of books, further issues which may affect these results, along with the aforementioned technical difficulties. These considerations also lead us to think that a discipline-focused study could shed more light on the validity of loans as a criterion for selecting monographs in selective indexes than could a broad study.

This study also confirms the need for facilities to aggregate different editions and translations of a certain book under one record in order to indicate the actual relevance of the overall work. It could be a future task for academic libraries to provide usage data, especially concerning book loans for regular systematic analyses, as they still constitute an important aspect of the scholarly communication process, though rarely recognized by bibliometric studies to date.

**Acknowledgments**

We would like to thank the Library of the University of Granada for facilitating the provision of the data for the purposes of this study. Nicolás Robinson-García is currently supported by an FPU grant from the Ministerio de Economía y Competitividad of the Spanish Government.